\documentclass[namedreferences]{kluwer}
\usepackage[]{graphicx}

\newcommand{\etal}{et~al.}

\newcommand{\Msun}{\ensuremath{\mathrm{M_{\odot}}}}

\begin{document}

\begin{article}
\begin{opening}         

\title{ARE ALL HOT SUBDWARF STARS\\ IN CLOSE BINARIES?}

\author{RICHARD~A. \surname{WADE}\email{wade@astro.psu.edu}}  
\author{M.~A. \surname{STARK}\email{stark@astro.psu.edu}}  
\institute{Pennsylvania State University}


\runningauthor{WADE and STARK}
\runningtitle{ARE ALL HOT SUBDWARF STARS IN CLOSE BINARIES?}

\date{June, 2003}

\begin{abstract}
 We discuss whether the hypothesis that ``all (or most) subdwarfs are
in close binaries'' is supported by the frequently reported
observations of photometrically or spectroscopically composite
character of many hot subdwarf stars.  By way of a possible
counter-argument, we focus on resolved companions (optical pairs) of
hot subdwarf stars.  On a statistical basis, many of these are
physically associated with the hot subdwarfs, i.e. are common proper
motion pairs.  These resolved pairs make a several percent
contribution to the catalog of hot subdwarf stars per decade in
projected separation.  If they are just the relatively wide members of
a binary population similar to the local G-dwarf binary population
\cite{DM91}, which has a very wide distribution of
orbital separations, then many of the unresolved but composite hot
subdwarf binaries may not be ``close'' in the astrophysical sense. In
that case, binary channels for hot subdwarf formation may be less
important than thought, or must involve companions (white dwarfs) that
do not result in a composite spectrum system.
\end{abstract}


\end{opening}       

An interesting number of spectroscopically identified hot sub\-dwarfs
\cite{KHD} appear to have resolved
companions.  It has been proposed that one or more binary
mechanisms are responsible for the formation of sdB stars in
particular (e.g., \opencite{Mengel}).  This would imply that most or
all of the sdB stars should have {\em close} companions.  Are
these sdBs with {\em distant} companions actually 
triple systems, so that they are also close binaries \cite{Maxted},
or instead do many field sdB stars {\em form without
binary interactions}?

We restrict attention here to the 959 sdB, sd, and sdB--O stars in
\inlinecite{KHD}, from which 31 ``best candidate" pairs were
selected, a rate of 3.2\%.  The median magnitude of the 31 sdB stars
is $V\sim15$, so the typical distance is $\sim$1 kpc, and the median
angular separation of the pairs, $\rho = 10''$, corresponds to
$\sim10^4$\,AU.

Using the surface density $n$ of field stars of comparable brightness,
one can estimate the probability,
$P(<\rho) = 1 - \exp (-\pi \rho^2 n)$, that a field star would appear
within angular distance $\rho$ by chance.  Selection by eye seems to
have found candidate pairs which have $P(<\rho) \sim$~2--3\% in the
mean ($\sim$1/2 of the candidate sample has $P(<\rho) < 1$\%).  We
thus expect a significant number of the sample of 31 candidate pairs
to be field contamination.  Nevertheless we claim that there is {\em a
significant excess over chance} (about 2--3\,$\sigma$) of these close
but resolved pairs, i.e., some of these pairs are physical binary
stars.  We intend to improve upon this first selection by eye to
produce a sample derived using the $P(<\rho)$ criterion directly.

Good photometric data in the visible are not available for most of the
candidate pairs, either resolved or in combined light.  The USNO-A2
catalog \cite{Monet} gives separate entries for the subdwarf
and its companion in 15 cases, with plausible photometry in ``blue''
($b$) and ``red'' ($r$) passbands.  The USNO-A2 magnitudes or colors
of an individual object are dubious, but the magnitude differences and
the color differences should be valid.  A majority of candidate
companions are brighter than the subdwarf in the ``$r$'' band. The
typical candidate companion is $\sim1$ mag redder (``$b-r$'') than its
hot subdwarf.  Six candidates (of 31) have $P(<\rho) < 2$\% and also
have USNO-A2 photometry. We assume that $r$ = Johnson $R$ and $b-r$ =
Johnson $B-R$, and we adopt $M_R = +4.5$ and $B-R = -0.3$ for a
typical sdB star.  This allows $r$ and $b-r$ data for the
candidate companions to be plotted in a CMD along with a main sequence
locus.  Although the sample size is small, the CMD is not inconsistent
with the notion that the resolved companion stars to field sdB stars
are largely main sequence G stars.

Better magnitude and color data are available for many of the
candidate pairs from the Two-Micron All Sky Survey (2MASS).  We
constructed a CMD assuming $M_J({\rm sd}) = 5.1$ and using the
observed $J_{\rm comp} - J_{\rm sd}$ and $(J-K_S)_{\rm comp}$ from
2MASS.  This diagram is not inconsistent with a main sequence nature
of the companion stars.  From color alone, F, G, and K
stars are indicated as the dominant companion type.

\begin{figure}
\centerline{\rotatebox{-90}{\includegraphics[width=18pc]{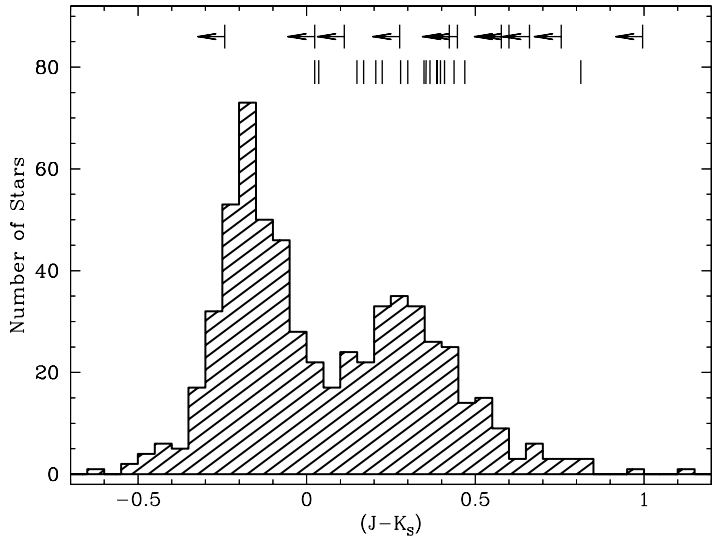}}}
 \caption{A histogram of $(J-K_S)$ for 612 (unresolved) sdB stars.
Tick marks and left--pointing arrows (upper limits) show the blended
colors of the candidate pairs.}  \label{JKhist}
\end{figure}

Figure~{\ref{JKhist}} shows a histogram of $(J-K_S)$ for
(unresolved) sdB stars (see Stark, Wade \& Berriman, this
volume), showing ``blue" (single) and
``red" (composite) cases.  Tick marks and
left--pointing arrows (upper limits) show the {\em blended}
$(J-K_S)$ colors of the candidate pairs, calculated from separate
2MASS measures of both components where possible.
Within sampling errors and bearing in mind the
contamination by field stars,
the $(J-K_S)$ distribution of the blended colors of the resolved
pairs is indistinguishable from that of the unresolved ``red"
(composite) subdwarfs.  This finding leads by another path to the
conjecture that the resolved pairs and the unresolved composites
are part of the same distribution: {\em many of the candidate resolved pairs
are the outer tail of a distribution of separations of physical
binaries}.

If the progenitors of sdB stars are solar--type stars (old disk
population), we expect them to show the binary population properties
of F--G dwarfs in the solar neighborhood, as summarized, e.g., by
\opencite{DM91} (DM91): roughly half of F--G ``primaries" will
have one or more companions; the distribution in $\log P$ is broad and
well--approximated by a Gaussian with mean period 180 years and
dispersion in $\log P$ of 2.3 decades.  For binaries with $M_1 + M_2 =
1\,\Msun$, the mean semimajor axis is $\bar{a} \approx 30$~AU, with a
dispersion in $\log a$ of 1.5 decades.

For a population of sdB stars with DM91--type companions at a typical
distance of 1 kpc, about 6\% would be resolvable with $\rho$ in the
4--40$''$ range.  We estimate that 1/3 to 1/2 of these would be
F--G--K companions ($\left|\Delta r\right| \leq 3$\,mag).  For 959 sdB
primaries, we thus expect about a dozen resolved pairs, in the range
of $\rho$ and $\Delta r$ to which the selection from Digitized Sky
Survey (DSS) images was sensitive.  The number of candidates found is
not inconsistent with the expected number.  The contamination rate is
of similar size, so the size of the observed statistical excess of
true binaries is not yet well established. 

We must validate that the candidate pairs are true c.p.m.\ systems on
a case--by--case basis.  Data allowing this validation are at present
fragmentary.

Using the DSS, the eye is only able to notice candidate pairs that
have $\rho >$~3--4$''$ typically.  Anecdotal, but strong,
circumstantial evidence exists that resolved, genuine sdB+ cool
companion pairs exist at smaller $\rho$, where field contamination is
significantly less likely and a DM91 scenario predicts more binaries
will be found.  Reports of smaller--separated pairs include \inlinecite{Thejll}, 
\inlinecite{Heber}, M.\, A.\, Stark (PG 1629+081
resolved), and R.\, A.\, Saffer (GD 108 ``barely resolved'').  Thus it
is clearly possible to search closer than the DSS allows and find
additional resolved binaries. A survey of $\sim 10^3$ sdB stars in the
0.5 to 2.0$''$ range could yield negligible field contamination (1--2
stars in 1000) and find $\sim$5\% of the sdBs with companions in this
range of $\rho$, if the pairs are of the DM91 type.

A systematic review of DSS images of sdB stars has found numerous
candidate sdB+companion pairs in the 4 to 40$''$ range of
separation. There seems to be a statistical excess of angularly close
but resolvable systems, corresponding to true wide binaries.  These
resolved pairs make a several percent contribution to the catalog of
hot subdwarf stars per decade in projected separation.  Evidence from
other studies supports the conjecture that most or many of the
``composite spectrum" sdB stars are merely binaries drawn from the
expected DM91--type distribution that solar--type stars should have.
Further ``close-in" studies, if carefully designed, can confirm or
refute the DM91 conjecture.

These wide and {\em non-interacting} companions, once individually
validated, may be exploited to ascertain the distance to the sdB
primary.  Also, the DM91 distribution, if validated, predicts that
only about 1/5 of the companions noted spectroscopically or
photometrically will be closer than $\sim 2$~AU and thus likely to
interact or have interacted with the sdB star.  Then {\em the majority
of cool companions that we actually identify are not participants in
close binary evolution}, and some mechanism not involving these
companions must be invoked to provide the right amount of mass loss on
the RGB and place the sdB stars where they are on the EHB. Whether
that mechanism involves a single star or a multiple--star system
(sdB+WD\,?)  cannot be established from these distant cool companions.

This research has made use of: 
the data products from the Two Micron All
Sky Survey;
the NASA/IPAC Infrared Science Archive;
the SIMBAD data\-base;
and the Digitized Sky Surveys.
This research has been supported in part by: 
NASA grant NAG5-9586, NASA GSRP grant NGT5-50399,  
and a NASA Space Grant Fellowship.

\end{article}
\end{document}